\documentclass[%
preprint%
,secnumarabic%
,tightenlines%
,amssymb, amsmath,nobibnotes, aps,nofootinbib,showpacs,]{revtex4}
\usepackage{color}
\usepackage{float}
\usepackage{amsmath,slashed,amssymb,epsf,epsfig,amsthm,bm,graphicx}
\definecolor{darkgreen}{rgb}{0,0.35,0}

\begin{document}

\title{Revisiting the gauge fields of strained graphene}
\author{Alfredo Iorio${}^1$}\thanks{E-mail: alfredo.iorio@mff.cuni.cz}
\author{Pablo Pais${}^{1,2,3}$}\thanks{E-mail: pais@cecs.cl}
\affiliation{$^1$ Faculty of Mathematics and Physics, Charles University - V Hole\v{s}ovi\v{c}k\'ach 2, 18000 Prague 8, Czech Republic}
\affiliation{$^2$ Centro de Estudios Cient\'{\i}ficos (CECs) \\ Av. Arturo Prat 514, Valdivia, Chile}
\affiliation{$^3$ Physique Th\'{e}orique et Math\'{e}matique, Universit\'{e} Libre de Bruxelles and International Solvay Institutes, Campus Plaine C.P.~231, B-1050 Bruxelles, Belgium}


\def\be{\begin{equation}}
\def\ee{\end{equation}}
\def\al{\alpha}
\def\bea{\begin{eqnarray}}
\def\eea{\end{eqnarray}}

\begin{abstract}
We show that, when graphene is only subject to strain, the spin connection gauge field that arises plays no measurable role, but when intrinsic curvature is present and strain is small, spin connection dictates most the physics. We do so by showing that the Weyl field associated with strain is a pure gauge field and no constraint on the $(2+1)$-dimensional spacetime appears. On the other hand, for constant intrinsic curvature that also gives a pure-gauge Weyl field, we find a classical manifestation of a quantum Weyl anomaly, descending from a constrained spacetime. We are in the position to do this because we find the equations that the conformal factor in $(2+1)$-dimensions has to satisfy, that is a nontrivial generalization to $(2+1)$-dimensions of the classic Liouville equation of differential geometry of surfaces. Finally, we comment on the peculiarities of the only gauge field that can describe strain, that is the well known {\it pseudogauge field} $A_1 \sim u_{11} - u_{22}$ and $A_2 \sim u_{12}$, and conclude by offering some scenarios of fundamental physics that this peculiar field could help to realize.
\end{abstract}



\maketitle

\section{Introduction}

Graphene is a very promising table-top laboratory to indirectly probe some of the fundamental mysteries of Nature \cite{iorio}. The low energy regime of its $\pi$ electrons is very well described by an effective theory that shares many of the features of a massless Dirac quantum field theory (QFT) in the presence of a background spacetime.

In order for graphene to keep its promises, we need to have full control of what sort of fields are there and what they represent in a field theory language. In the vast literature on the gauge fields of graphene (see \cite{gauge_fields_graphene, pacoreview2009, pseudomagexper}, and \cite{latestreviewmadrid} for a recent review), there are a variety of proposals, sometimes practically valuable for the applications to condensed matter physics, but most of the time unsatisfactory for probing fundamental physics. The landscape of proposals ranges from $SU(2)$ monopolelike gauge fields in the case of graphene membranes with intrinsic curvature (the inflated graphene buckyballs of \cite{vozmediano2}) to a concurrence of a spin-connection field and a $U(1)$ field, in the case of purely strained graphene \cite{dejuan-sturla-vozmediano} (although sometimes non-Abelian fields are evoked in this case as well \cite{dejuannonabelian}). Even in the simplest case, that is purely strained graphene, there is some confusion: does the spin connection arising from straining graphene give physical effects or not? And, what is the interpretation of the $U(1)$ field from a fundamental point of view?

In this paper, we try to clarify part of this confusion, having in mind to come back to the full scenario of gauge fields in graphene in a forthcoming work. Our principal conclusions are the following: (1) When only strain is present, spin connection plays no role; only the $U(1)$ pseudogauge field gives physical effects. (2) When intrinsic curvature is present, spin connection plays a crucial role, to the point that, for small strain, it is spin connection that dictates most of the physics. In addition, we also clarify why some confusion might arise in relation to the non-Abelian structure (that is the Lorentz group) behind the Weyl (Abelian) pure gauge field \cite{originalWeyl, lor}.

To do that, we proceed as follows. We start off with the fully relativistic approach, and proceed by seeing that the Weyl field associated to strain (that is what the spin connection reduces to in this case) is a pure-gauge field, that can never give a physical effect, unless non-trivial topology is involved. Since we are in a pure strain configuration (elastic deformations), and we only focus on local properties (our samples here have infinite size), non-trivial topology of the Weyl field is excluded {\it a priori}. Later we show, for the first time explicitly, that this fully relativistic approach is equivalent to the standard (non-fully relativistic, static Hamiltonian) approach. We do so by showing that all the terms one obtains through the latter method are there also in the former. Having the equivalence, this closes the debate on the physical effects of the spin-connection for pure strain.

On the other hand, a crucial issue arises for the meaning of Weyl symmetry itself. In fact, the Weyl field is a pure gauge \textit{also} when the membrane has nonzero constant (hence non-trivial) intrinsic curvature, and we know that physical effects emerge from there \cite{ioegae}. How can we reconcile this, on general grounds, with the previous result? We have solved the issue by first finding the equations that the conformal factor of the $(2+1)$-dimensional theory has to satisfy, in the case of zero curvature (pure strain) and in the case of constant curvature. This is interesting on its own right, as we have found a nontrivial generalization to $(2+1)$ dimensions of the classic Liouville equation of two-dimensional differential geometry of surfaces, and of $(1+1)$-dimensional conformal field theories (CFTs). We then realize that the key difference between the flat and curved case, although they both give standard Weyl invariance, lie in the absence or presence, respectively, of constraints on the $(2+1)$-dimensional spacetime. Therefore, in the curved case, the theory effectively lives in $(1+1)$ dimensions where Weyl symmetry is anomalously broken. This is an interesting classical manifestation of a quantum Weyl anomaly.

The paper flow is as follows: In Section~\ref{top-down} we recall the results of previous work that we need here, and set the notation. In Section~\ref{sigmaSigma} we find the connection between spatial and spatiotemporal conformal factors that allow us to conclude that the spin-connection cannot give physical effects for pure strain. We focus the last part of this section on the nonzero constant curvature case. We leave to Section~\ref{contact} to show explicitly how the standard Hamiltonian and strain tensor language can be recovered from the fully relativistic formalism. After this {\it top-down} approach, in the Section \ref{bottom-up}, we use the tight-binding condensed matter Hamiltonian description, and comment on the peculiarities of the only gauge field that can describe strain. This latter road, that we call {\it bottom-up}, is actually necessary because, this kind of gauge field could not be guessed within standard QFT in curved spacetime. In the concluding Section we also offer some scenarios of fundamental physics that this peculiar gauge field could help to realize.

\section{Spin-connection and Weyl gauge field for strained graphene} \label{top-down}

As is well known (see, e.g., \cite{pacoreview2009}), the physics of the large wavelength $\pi$ electrons of graphene, at half-filling, can be efficiently encoded within the Dirac massless two-dimensional {\it Hamiltonian}. In our approach, we include time to make it fully relativistic, although with the speed of light $c$ traded for the Fermi velocity $v_F$ (see, e.g., \cite{originalWeyl}); hence, we start from the $(2+1)$-dimensional action
\be \label{flatactionWeyl2+1}
{\cal A} = i \hbar v_F \int d^3 q \bar{\psi} \gamma^a \partial_a \psi \;,
\ee
where $q^a = (t,x,y)$ are the flat spacetime coordinates, $\gamma^{a}$ are the usual Dirac matrices in three dimensions, and we expand around only one of the two Dirac points, as we shall focus on strain.

As shown in \cite{originalWeyl}, the metric
\begin{equation}\label{mainmetric}
g_{\mu \nu}  (q) = \left(\begin{array}{cc} 1 & 0  \quad 0 \\ \begin{array}{c} 0 \\ 0 \end{array} & - g_{i j} (x,y) \\ \end{array} \right)\;
\end{equation}
can also describe strain; hence, we shall use the customary Dirac action in that curvilinear background. As also pointed out in \cite{originalWeyl}, we recall that although the system is $(2+1)$ dimensional, the Riemann tensor ${R^\lambda}_{\mu \nu \rho}$, $\lambda, ... = \underline{0}, \underline{1},\underline{2}$, has only one independent component, proportional to the Gaussian curvature, $\cal K$. Furthermore, surfaces of zero or constant $\cal K$, make the metric \eqref{mainmetric} flat or conformally flat, respectively, and both cases can be treated at once within a formalism that uses
\be\label{metricCF2+1}
g_{\mu \nu} (Q)  = e^{2 \Sigma(Q)} \eta_{\mu \nu} \;,
\ee
where $\eta_{\mu \nu} = {\rm diag} (1, -1, -1)$, and the information about the metric being flat or not is encoded in the conformal factor $\Sigma$.
The coordinates where the metric can be explicitly written in a conformally flat fashion,
\be \label{CFcoordinates}
Q^\mu \equiv (T,X,Y) \;,
\ee
are, in general, different from the original coordinates $q^\mu$.

Therefore, the natural candidate action to describe strained graphene is (for a while we shall use $\hbar = 1 = v_F$)
\be \label{actionWeyl2+1}
{\cal A} = i \int d^3 Q \sqrt{g} \bar{\psi} E_a^\mu \gamma^a (\partial_\mu + \Omega_\mu) \psi \;,
\ee
where $\Omega_\mu = \frac{1}{2} \omega_\mu^{\; a b} J_{a b}$, with $J_{a b} = \frac{1}{4} [\gamma_a, \gamma_b]$, the Lorentz generators, $E_a^\mu$ is the inverse of the three dimensional vielbein $e^{a}_{\mu}$ (the dreibein), $\omega_\mu^{\; a b}$ is the spin connection, and being in $(2+1)$ dimensions, we can write the spin connection in the one-index notation
\be \label{omega2+1}
\omega^a_\mu = \frac{1}{2} \epsilon^{a b c} \omega_{\mu \; b c} .
\ee
Here, $a$ is the non-Abelian index of the local Lorentz group SO(2,1), and $\mu$ is the vector index on the spacetime manifold. In this formalism, the index $a$, a tangent spacetime index, plays the role of an internal/gauge index.

The metric \eqref{mainmetric} can always be written in more suitable coordinates $\tilde{q}^\mu = (t, \tilde{x}, \tilde{y})$, where $t$ is the same time for both coordinate systems, and $(\tilde{x}, \tilde{y})$ are the {\it spatial isothermal} coordinates of the surface
\begin{equation}\label{metricisothermal2+1}
g_{\mu \nu}  (\tilde{q}) = \frac{\partial q^\lambda}{\partial \tilde{q}^\mu}
\frac{\partial q^\kappa}{\partial \tilde{q}^\nu} g_{\lambda \kappa} (q) =
\left(\begin{array}{ccc} 1 & 0  & 0 \\
0 & - e^{2 \sigma(\tilde{x}, \tilde{y})} & 0 \\
0 & 0 & - e^{2 \sigma(\tilde{x}, \tilde{y})}\end{array} \right) \;.
\end{equation}

The single scalar function $\sigma$ identifies the surface/graphene membrane. Other isothermal coordinates can be found, say $\tilde{\tilde x}, \tilde{\tilde y}$, but the function identifying the surface is always the same\footnote{As an example of formula \eqref{intrinsicsigma}, the metric of a sphere of radius $r = 1$ could be written as $ds^{2}=e^{2\sigma}\left(d\tilde{x}^{2}+d\tilde{y}^{2}\right)$ with conformal factor $\sigma (\tilde{x}, \tilde{y}) = \ln (1/\cosh \tilde{y})$ being $\tilde{x}\in[0,2\pi]$ and $\tilde{y}\in(-\infty,+\infty)$. On the other hand, using the coordinates $\tilde{\tilde y} = \begin{cases} \arcsin (1/\cosh \tilde{y}), & \mbox{if } \tilde{y}\geq 0 \\ \pi - \arcsin (1/\cosh \tilde{y}), & \mbox{if } \tilde{y}< 0 \end{cases}$, and $\tilde{\tilde x} = \tilde{x}$, $\sigma(\tilde{\tilde x}, \tilde{\tilde y}) = \ln \sin \tilde{\tilde y}$, with $\tilde{\tilde y} \in (0, \pi)$, we recover the standard metric of the sphere $ds^{2}=\sin^{2}\tilde{\tilde{y}} d\tilde{\tilde{x}}^{2}+d\tilde{\tilde{y}}^{2}$.}:
\be \label{intrinsicsigma}
\sigma(\tilde{\tilde x} (\tilde{x}, \tilde{y}), \tilde{\tilde y} (\tilde{x}, \tilde{y})) = \sigma (\tilde{x}, \tilde{y}) \;.
\ee

\section{Connecting space and spacetime conformal factors} \label{sigmaSigma}

When the surface described by $\sigma$ has zero or constant curvature, the two metrics \eqref{metricCF2+1} and \eqref{metricisothermal2+1} both describe the same spacetime, although with different coordinates,
\be
g_{\mu \nu} (Q) = e^{2 \Sigma(Q)} \eta_{\mu \nu} = \frac{\partial \tilde{q}^\lambda}{\partial Q^\mu}
\frac{\partial \tilde{q}^\kappa}{\partial Q^\nu} g_{\lambda \kappa} (\tilde{q}) \;;
\ee
hence, their Gaussian curvature cannot differ. Due to the simple structure of the metric in the two coordinate frames, it is easy to compute the Gaussian curvature
\bea
{\cal K}  & = & -  e^{- 2 \sigma} [\nabla^2 \sigma] \\
& = & e^{- 2 \Sigma} \left[ 2 \Box \Sigma + (\partial_a \Sigma) (\partial^a \Sigma) \right]
\eea
where $\nabla^2 = \partial^2_{\tilde{x}} + \partial^2_{\tilde{y}}$, and $\Box = \partial^2_T - \partial^2_X - \partial^2_Y$. Clearly, the two conformal factors are related, $\Sigma (\sigma)$: if we know $Q^\mu(\tilde{q})$, we can write $\Sigma(\tilde{x}, \tilde{y}) = \Sigma (T(\tilde{x}, \tilde{y}), X(\tilde{x}, \tilde{y}), Y(\tilde{x}, \tilde{y}))$, and then knowing $\sigma(\tilde{x}, \tilde{y})$, we obtain $\Sigma (\sigma)$. Nonetheless, we have the general equations that $\Sigma$ has to satisfy for the two cases, for ${\cal K}=0$,
\be\label{eqforSigmaflat}
\Box \Sigma = - \frac{1}{2} (\partial_a \Sigma) (\partial^a \Sigma) \;,
\ee
which corresponds to $\sigma$ harmonic functions (i.e., solutions of $\nabla^2 \sigma = 0$), and for $\cal{K} =$constant
\be\label{eqforSigmacurved}
\Box \Sigma = - \frac{1}{2} (\partial_a \Sigma) (\partial^a \Sigma) + \frac{1}{2} {\cal K} e^{ 2 \Sigma} \;,
\ee
which corresponds to $\sigma$ Liouville functions (i.e., solutions of $\nabla^2 \sigma = - {\cal K} e^{2 \sigma}$).

Let us focus on the flat case which, since it corresponds to pure strained graphene, is the one of interest here. In this case, besides the obvious constant solution of \eqref{eqforSigmaflat}, $\Sigma_{flat} = C$, we also have
\be\label{Sigmaflat}
\Sigma_{flat} = - \ln (T^2 - X^2 - Y^2) + C \;.
\ee
The constant $C$ could be set to zero, but we shall keep it, to later compare with the curved case. For the conformal factor \eqref{eqforSigmaflat}, the metric \eqref{metricCF2+1} reads
\begin{equation}
g_{\mu \nu}  (Q) = e^{2 \Sigma_{flat}} \eta_{\mu \nu}
= \frac{e^{2 C}}{(T^2 - X^2 - Y^2)^2}
\left(\begin{array}{ccc} 1 & 0  & 0 \\ 0 & - 1 & 0 \\ 0 & 0 & - 1 \end{array} \right) \;.
\end{equation}
Hence, nothing constrains the norm of vectors just as for the Minkowski case
\be\label{normflat}
\| Q \|^2 = g_{\mu \nu} Q^\mu Q^\nu = \frac{e^{2C}}{ \eta_{\mu \nu} Q^\mu Q^\nu}
\ee
that means that $Q^\mu$ truly describes a flat three-dimensional spacetime. In order to see explicitly that \eqref{normflat} is in fact the Minkowski spacetime, take the following change of coordinates:
\begin{eqnarray}
t&=&\frac{e^{C}T}{T^{2}-X^{2}-Y^{2}}, \nonumber \\
x&=&\frac{e^{C}X}{T^{2}-X^{2}-Y^{2}}, \label{change_coordinates} \\
y&=&\frac{e^{C}Y}{T^{2}-X^{2}-Y^{2}}. \nonumber \\
\end{eqnarray}
On these coordinates the line element is $ds^{2}=dt^{2}-dx^{2}-dy^{2}$, showing that the singularities appearing in the metric \eqref{normflat} are simply coordinates singularities due to our choice of nonstandard coordinates\footnote{To see whether one has a true or a coordinate singularity is, in general, not an easy task. On this, see, e.g., \cite{wald}.}.
As a result of that, when we use $\Sigma_{flat}$ in the spin connection of \eqref{actionWeyl2+1}
\be\label{aomegacf}
    {\omega_\mu}_{a b} = \delta^c_\mu (\eta_{c a} \delta^\nu_b - \eta_{c b} \delta^\nu_a) \Sigma_\nu \;,
\ee
we can safely use $\delta_a^\mu$ as a proper dreibein, because it indeed connects the tangent space with a flat manifold. Here $\Sigma_\mu = \partial_\mu \Sigma$, and $\Sigma_a = \partial_a \Sigma$, and we used the result that in three dimensions $\gamma^a J_{a b} = \gamma_b$.

\subsection{Zero curvature: no physical effects of strain through the spin-connection} \label{zeroomega}

Therefore the action \eqref{actionWeyl2+1} for the metric \eqref{metricCF2+1} is
\begin{equation}\label{afinal}
{\cal A} = i \int d^3 Q \; e^{2 \Sigma} \; \bar{\psi} \gamma^a (\partial_a + \Sigma_a) \psi \;,
\end{equation}
and when the Dirac field is properly transformed
\be\label{psiWeyl}
\psi = e^{- \Sigma(Q)} \psi' \;,
\ee
where $\psi'$ refers to the Minkowskian flat spacetime $\eta_{\mu \nu}$ [see \eqref{metricCF2+1}], the action \eqref{actionWeyl2+1} is simply
\be \label{flataction2+1}
{\cal A} = i \int d^3 Q \bar{\psi}' \gamma^a \partial_a \psi' \;,
\ee
which refers to the background metric $\eta_{\mu \nu}$, which in turn is the unstrained situation. Strain is gone altogether. It has no physical effects.

Another way to see this, is to stop at the action \eqref{afinal}, i.e., {\it before} implementing the transformation of the spinor as in \eqref{psiWeyl}, and consider the gauge field $\Sigma_a$. This gauge field is itself a pure derivative, hence it cannot produce any physical effect through its field strength
\be
F_{a b} = \partial_a \Sigma_b - \partial_b \Sigma_a = (\partial_a \partial_b - \partial_b \partial_a) \Sigma = 0 \;,
\ee
as can be seen also by explicitly computing $\vec{B}^{\Sigma_{flat}} \equiv \vec{\nabla} \times \vec{\Sigma}_{flat}$ and $\vec{E}^{\Sigma_{flat}} \equiv - \vec{\nabla} \Sigma^0 + \partial_T \vec{\Sigma}_{flat}$ with $\Sigma_{flat}$ in \eqref{Sigmaflat}. The result is zero $\vec{B}^{\Sigma_{flat}} = 0 = \vec{E}^{\Sigma_{flat}}$.

Therefore, from here, we see that the very well-known pseudomagnetic field (and, for what matters, even a putative pseudoelectric field) induced by pure strain, cannot be accounted for by the spin-connection/Weyl pure-gauge field.

Let us now comment on $\Sigma_a$. As seen, this is a pure gauge field associated to the local Weyl transformations \eqref{metricCF2+1} and \eqref{psiWeyl}. Indeed, the Weyl field transforms as
\be
W_\mu \to W_\mu - \partial_\mu \Sigma \;.
\ee
The reason why we do not have here the full Weyl gauge field, $W_\mu$, but only its pure gauge part, is due to the local Weyl invariance of \eqref{actionWeyl2+1}; see \cite{originalWeyl}. Like any other Weyl field, $\Sigma_a$ is an Abelian gauge field. Abelian gauge fields are those routinely used in the graphene literature, \cite{pacoreview2009} and \cite{dejuan-sturla-vozmediano}. On the other hand, $\Sigma_a$ carries information on the non-Abelian structure of the local Lorentz group that is encoded in the spin connection. This information is in the tangent space index ``$a$'' of $\Sigma_a$; see \eqref{omega2+1}, \eqref{aomegacf} and discussion in between. Indeed, (local) Weyl transformations, in general, rephrase spacetime scaling as an internal transformation. Non-Abelian gauge fields have also appeared in various discussions on the gauge field approach to strained graphene \cite{dejuannonabelian}.

The above mentioned properties are common to the full Weyl gauge field $W_\mu$; hence, they hold also for theories that do not have local Weyl invariance. The extra property of $\Sigma_a$ is that it is $\partial_a \Sigma$; i.e. the true degree of freedom is just one scalar, $\Sigma$, the one related to the two-dimensional spatial phonon, $\sigma$, of the graphene membrane, as shown before.

These facts are fully transparent in the coordinates $Q^\mu$ in \eqref{CFcoordinates}. When the coordinates are different, the three aspects of this gauge field---(i) scalar nature, (ii) abelian field, and (iii) non-Abelian Lorentz structure---get mixed together, and they may appear, in the standard coordinates/frames used in graphene, as originating from different gauge fields, as we show in Section~\ref{contact}.

\subsection{Nonzero curvature: the classical manifestation of the quantum Weyl anomaly} \label{weylanomaly}

Apparently, all seems clear: when $\Sigma = \Sigma_{flat}$, which should describe pure strain, no physical effects can be described by the QFT in the curved spacetime approach. Nonetheless, we also saw that when the Gaussian curvature of the membrane is constant, a similar procedure could be applied; hence, this seems to lead to conclude that also in that case as well, there is no physical effect. However, as we shall now show, this is not the case.

When one solves \eqref{eqforSigmacurved}, e.g., for negative curvature, ${\cal K} = - r^{-2}$, one obtains
\be\label{Sigmacurved}
\Sigma_{curved} = - \frac{1}{2} \ln (T^2 - X^2 - Y^2) + \ln r \;,
\ee
and evidently the associated $\vec{B}^{\Sigma_{curved}}$ and $\vec{E}^{\Sigma_{curved}}$ are zero.

\begin{figure}[H]
\begin{center}
\includegraphics[width=0.4\textwidth,angle=0]{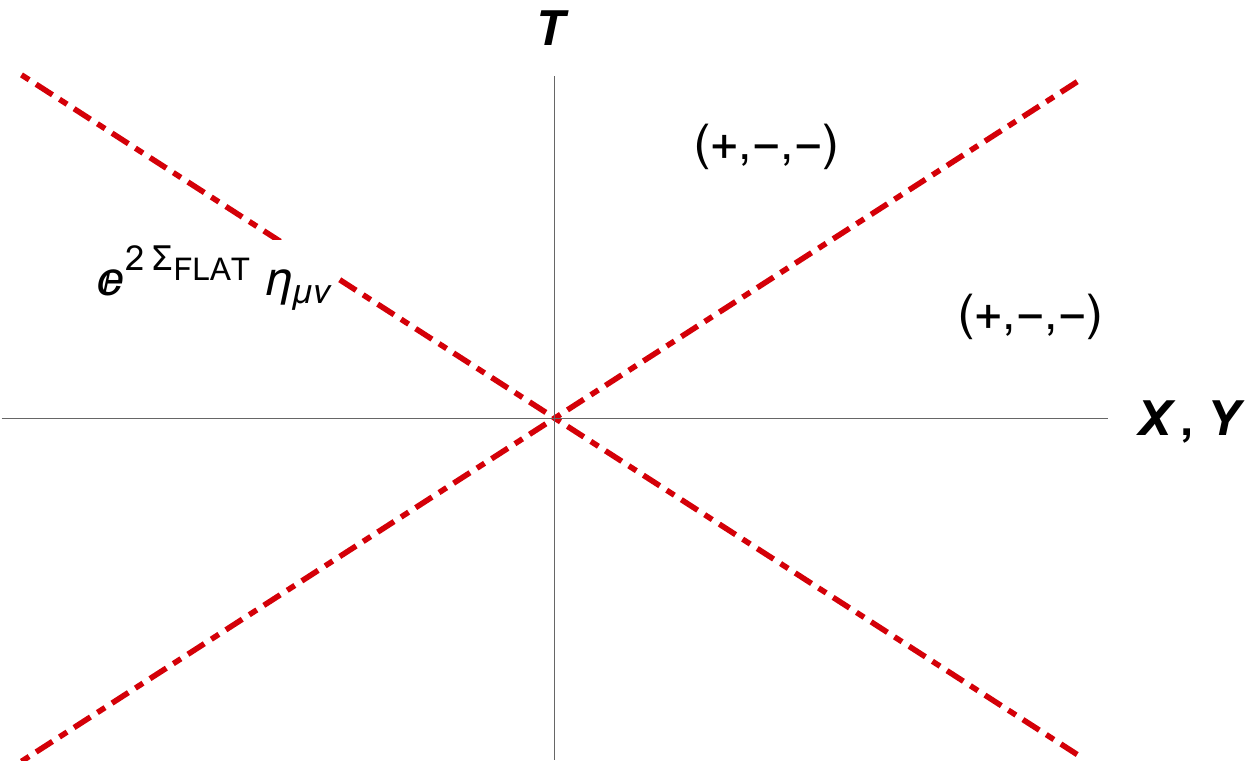}
\includegraphics[width=0.4\textwidth,angle=0]{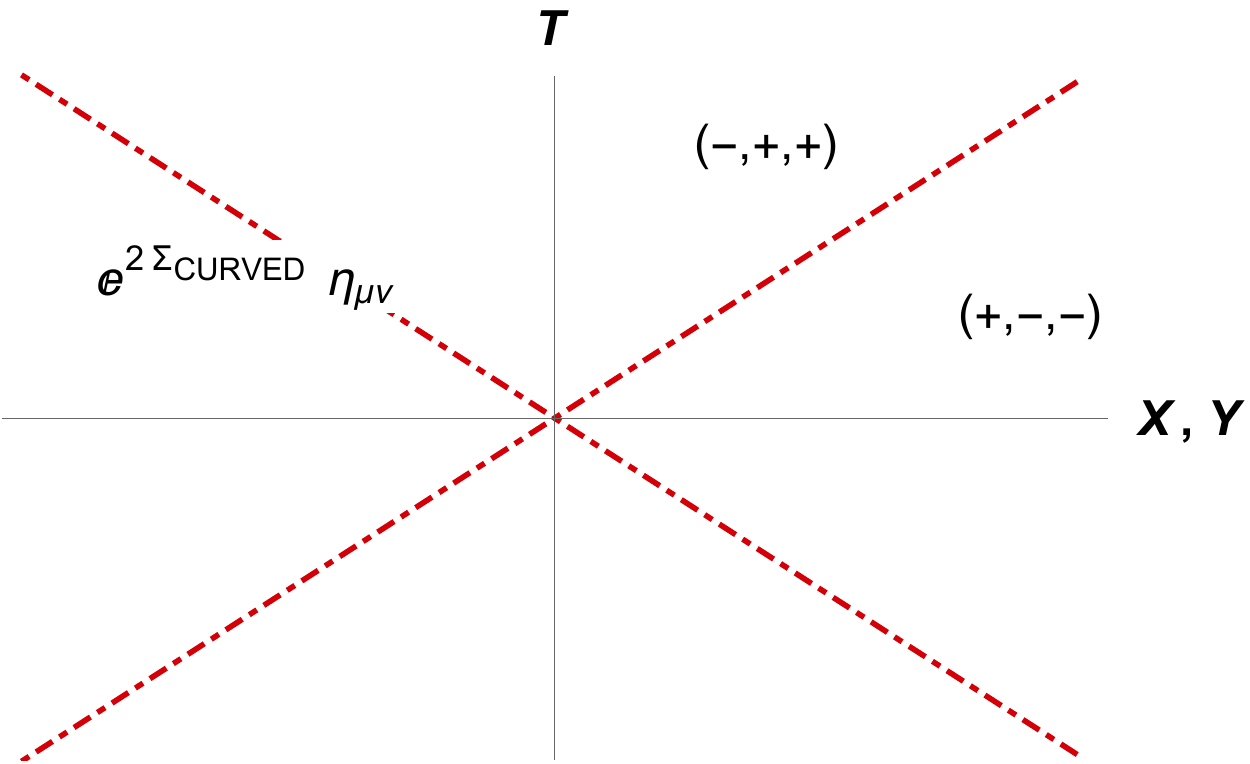}
\end{center}
\caption[3pt]{{\protect\small {For $\Sigma_{flat}$ (figure on the left) the spacetime is essentially the same as Minkowski. Modulo an inessential singularity at the light cone, the causal structures are identical; hence, the spacetime is three-dimensional and flat all the way. In the case corresponding to $\Sigma_{curved}$ (figure on the right), we see that there is a change of signature crossing the light cone, that signals a behavior similar to a black hole horizon. In this latter case, the true spacetime is one dimension smaller, due to the constraint \eqref{normcurved} that coordinates need to satisfy when curvature is present.}}}%
\label{newspacetimes}%
\end{figure}

Nonetheless, this time the metric \eqref{metricCF2+1} reads
\begin{equation}
g_{\mu \nu}  (Q) = e^{2 \Sigma_{curved}} \eta_{\mu \nu}
= \frac{r^2}{T^2 - X^2 - Y^2}
\left(\begin{array}{ccc} 1 & 0  & 0 \\ 0 & - 1 & 0 \\ 0 & 0 & - 1 \end{array} \right) \;,
\end{equation}
and the light-cone becomes essentially singular due to the change of signature, typical of horizons in black hole spacetimes. Hence, the light-cone is a proper Killing horizon, as shown already when considering Hawking effects reproduced on graphene; see, e.g., \cite{ioegae}. Let us now see what happens to the length of a position vector,
\be\label{normcurved}
\| Q \|^2 = g_{\mu \nu} Q^\mu Q^\nu = r^2 > 0 \;.
\ee
That is a dramatic difference with the previous case \eqref{normflat} (amusingly, the difference comes about because of the different multiplicative constant factors: $- 1$ for the flat vs $- 1/2$ for the curved case). In the curved case the coordinates obey a constraint; hence, the effective theory lives in one dimension less, with the coordinates that we call $P^\alpha$, $\alpha = \underline{0},\underline{1}$. The behavior of the flat and curved case is depicted in Fig. \ref{newspacetimes}. Furthermore, the $\delta_a^\mu$ in the expression for the spin-connection \eqref{aomegacf} is not the dreibein anymore: the manifold index $\mu$ now refers to a true curved space. All of this makes us conclude that the action that corresponds to the curved case (in the negative curvature case) is
\be
{\cal A}^{(2)}_{eff} \thickapprox i \int d^2 P \sqrt{g^{(2)}} \bar{\chi} \gamma^\alpha (\partial_\alpha + \Omega_\alpha) \chi \;,
\ee
where everything has been dimensionally reduced to the two dimensions and, in particular, the metric is the induced curved metric:
\be
g^{(2)}_{\alpha \beta} (P) = \eta_{a b} \frac{\partial Q^a}{\partial P^\alpha} \frac{\partial Q^b}{\partial P^\beta} \;.
\ee
As well known, this two-dimensional theory has a quantum Weyl anomaly. Hence, interestingly, through the classical constraints we have a manifestation of a quantum Weyl anomaly.

The Weyl (trace) anomaly is known to be in one-to-one correspondence with the Hawking radiation \cite{chr}. This is an interesting road to pursue for an alternative computation of the Hawking phenomenon on graphene as the one present in the literature \cite{ioegae}.

\section{Equivalence of the static Hamiltonian and the fully relativistic approaches}\label{contact}

To make contact with the literature focused on the phenomenology of graphene (see, e.g., \cite{latestreviewmadrid}) we need first to move from the action in the $Q^\mu$ coordinates to the Hamiltonian in the $\tilde{q}^\mu$ coordinates, keeping the curvature radius $r$ finite. For $\Sigma = - \frac{1}{2} \ln (T^2 - X^2 - Y^2) + \ln r = - t/r$, the $Q^\mu$ coordinates are
\bea
T & = & e^{t/r} \sqrt{e^{2 \sigma (\tilde{y})} + r^2} \label{coord1}\\
X & = & e^{t/r} e^{\sigma (\tilde{y})} \cos \tilde{x} \label{coord2} \\
Y & = & e^{t/r} e^{\sigma (\tilde{y})} \sin \tilde{x} \label{coord3}
\eea
To write the Hamiltonian in the $\tilde{q}$ coordinates, we have to consider the action \eqref{afinal}, Legendre-transform it
\be
\mathcal{A}=\int dtd\tilde{x}d\tilde{y}\mathcal{L}(\tilde{q})=\int dt d\tilde{x}d\tilde{y} \left[\left\|\frac{\partial Q}{\partial \tilde{q}}\right\|(\pi_{\psi}\partial_{T}\psi)(\tilde{q})-\mathcal{H}\right],
\ee
and follow the steps presented in the Appendix A. This gives, for $H=\int d\tilde{x}d\tilde{y} \mathcal{H}(\tilde{q})$,
\be \label{H2phiavf(x)}
H = - i \hbar \int d\tilde{x} d\tilde{y} \; \psi^\dag \left( \tau^i \left[ v_F(\sigma(\tilde{y}))\right]_i^{\; \; \tilde j} \; \partial_{\tilde j} +  v_F \phi + v_F \tau^i A_i \right) \psi
\ee
where $\tau^{i}$ are the Pauli matrices, we have re-introduced $\hbar$ and $v_F$, and
\be \label{phiexplicit}
\phi = \frac{1}{r} \; e^{2 \sigma} \sigma_{\tilde y}
\ee
\be \label{Aexplicit}
A_{1} = -\frac{1}{r} \; \frac{\; e^{3 \sigma} \; \sigma_{\tilde{y}}}{\sqrt{e^{2\sigma} + r^2}} \; \cos \tilde{x} \;, \;
A_{2} = -\frac{1}{r} \; \frac{\; e^{3 \sigma} \; \sigma_{\tilde{y}}}{\sqrt{e^{2\sigma} + r^2}} \; \sin \tilde{x} \;,
\ee
\be \label{fermiveltensor}
\left[ v_F(\sigma(\tilde{y}))\right]_i^{\; \; \; \tilde j} = v_F \; \left(\begin{array}{cc} v_{1 \tilde{x}} & v_{1 \tilde{y}} \\
v_{2 \tilde{x}} & v_{2 \tilde{y}} \\ \end{array} \right) \;,
\ee
with
\bea
v_{1 \tilde{x}} & = &  \frac{r \; e^{\sigma} \; \sigma_{\tilde{y}}}{\sqrt{e^{2\sigma} + r^2}} \; \sin \tilde{x} \\
v_{1 \tilde{y}} & = & -\frac{e^\sigma}{r} \; \sqrt{e^{2\sigma} + r^2} \; \cos \tilde{x} \\
v_{2 \tilde{x}} & = & -\frac{r \; e^{\sigma} \; \sigma_{\tilde{y}}}{\sqrt{e^{2\sigma} + r^2}} \; \cos \tilde{x} \\
v_{2 \tilde{y}} & = & -\frac{e^\sigma}{r} \; \sqrt{e^{2\sigma} + r^2} \; \sin \tilde{x}\;.
\eea

The computations here were carried on for the $\Sigma$ in \eqref{Sigmacurved}, for which we can present the coordinates \eqref{coord1}-\eqref{coord3}; hence, the expressions for $\phi$, $A_i$ and $\left[ v_F \right]_i^{\; \; \; \tilde j}$ depend of that choice. Nonetheless, even though the detailed expression of those quantities change for $\Sigma$ in \eqref{Sigmaflat}, the structure of the Hamiltonian \eqref{H2phiavf(x)} remains the same for the case of interest of pure strain.

We see here that, through this top-down method, we were able to reproduce all the terms except one that customarily appear in the literature of strained graphene; see, e.g., \cite{latestreviewmadrid}. The latter is the one gauge field that gives unambiguous physical effects, and that couples to the spinors with an imaginary factor (an instance that, on its own right, is an indication that such field cannot be a Weyl field, see, e.g., \cite{originalWeyl}). In the next Section we shall extensively comment on this field. Before moving to that, let us get one step closer to the language usually adopted in the graphene literature.

In fact, the former expressions are written in the language of conformal factors $\sigma$ for the membrane and in isothermal coordinates whereas the usual approach employs strain tensors $u_{i j} = 1/2 (\partial_i u_j + \partial_j u_i)$ (where, as customary, the $u_i$ measures the departure from the unstrained position), and Cartesian coordinates.

Although we started off with a fully relativistic formalism, due to the structure of the metric \eqref{mainmetric}, everything of the previous expressions necessarily depends only on the spatial coordinates. Henceforth, we can focus on the spatial metric only and make the customary ansatz that, in Cartesian coordinates $(x,y)$,
\be \label{guij}
g_{i j} (x,y) \simeq \delta_{i j} + 2 u_{i j} \;.
\ee
On the other hand, this metric is related through a coordinate change to the one in \eqref{metricisothermal2+1}
\be \label{legesonsigma}
g_{i j} (q) = \frac{\partial \tilde{q}^k}{\partial q^i} \frac{\partial \tilde{q}^l}{\partial q^j} \; \delta_{k l} e^{2 \sigma (\tilde{q})} = L_{i j} \; e^{2 \sigma (\tilde{q} (q))} \;.
\ee
This needs to be considered in its infinitesimal form, i.e., $\tilde{q}^i (q) \simeq q^i + \tilde{u}^i$, so that
$\partial \tilde{q}^i/ \partial q^j \simeq \delta^i_j + \partial_j \tilde{u}^i$, which gives, at first order,
$L_{i j} \simeq \delta_{i j} + 2 \tilde{u}_{i j}$; hence,
\be \label{gutildeij}
g_{i j} (x,y) \simeq \left( \delta_{i j} + 2 \tilde{u}_{i j} \right) (1 + 2 \sigma) \;.
\ee
Comparing the two expressions \eqref{guij} and \eqref{gutildeij}, we obtain the wanted link between conformal factor, Cartesian strain tensor, and isothermal strain tensor
\be
u_{i j} \equiv \sigma \delta_{i j} + \tilde{u}_{i j}  \;.
\ee

\section{The honeycomb structure gauge field} \label{bottom-up}

\begin{figure}
\begin{center}
\includegraphics[width=0.6\textwidth]{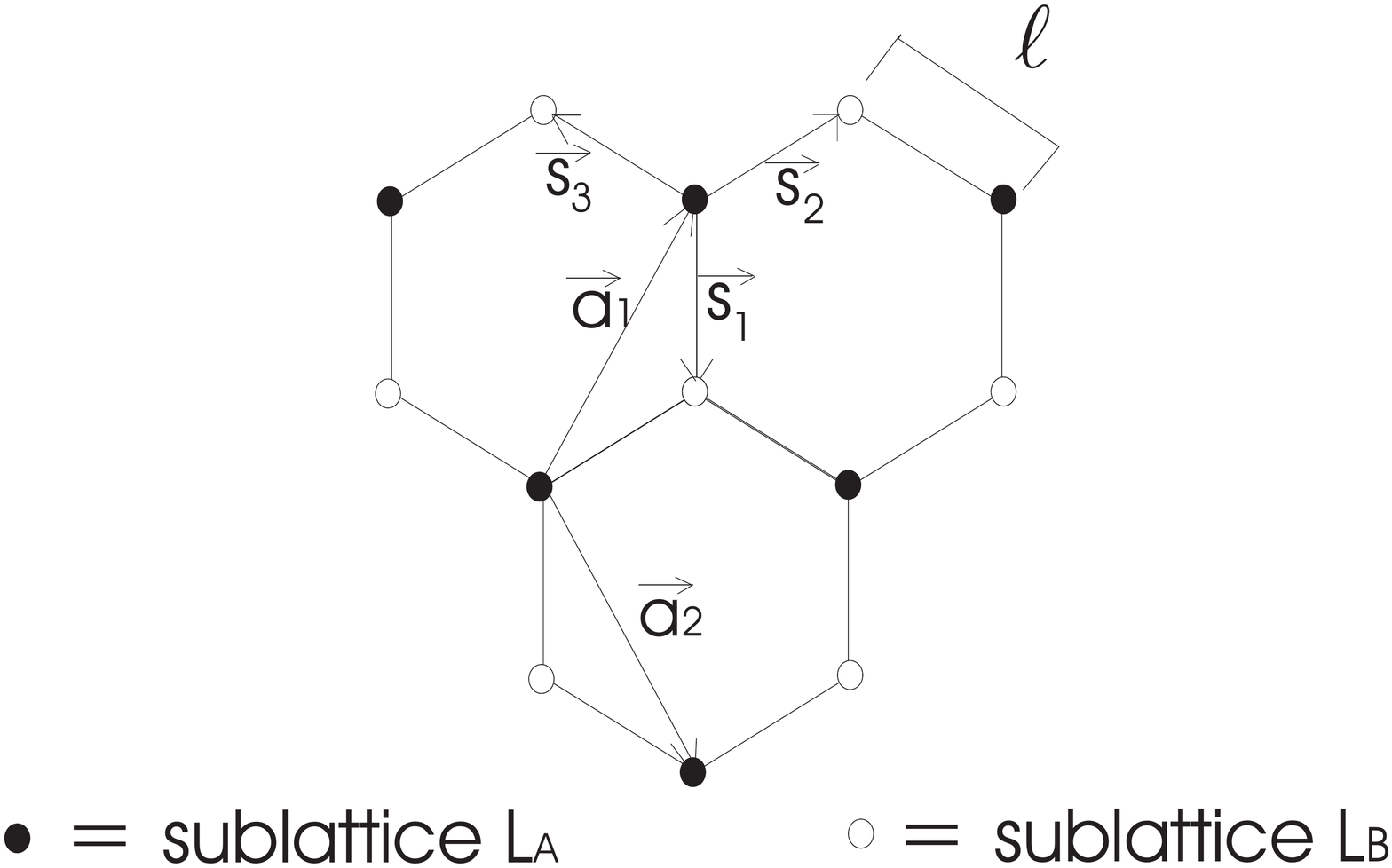}
\end{center}
\caption[3pt]{{\protect\small {The honeycomb lattice of graphene, and its two triangular sublattices, with our choice of the basis vectors,
$\vec{a}_1 = \frac{\ell}{2} (\sqrt{3}, 3)$, $\vec{a}_2 = \frac{\ell}{2} (\sqrt{3}, - 3)$, and
$\vec{s}_{1}=\ell(0,-1)$, $\vec{s}_{2}=\frac{\ell}{2}(\sqrt{3},1)$, $\vec{s}_{3}=\frac{\ell}{2}(-\sqrt{3},1)$, with $\ell$ the carbon-to-carbon distance.}}}%
\label{honeycomb}%
\end{figure}

Near the Dirac points (with our choice of the basis vectors in Fig.~\ref{honeycomb}, $\vec{k}_{\pm}^{D}=\left(\pm\frac{4\pi}{3\sqrt{3}\ell},0\right)$), we can describe the behavior of the $\pi$ electrons of graphene with the following Dirac Hamiltonian,
\begin{equation}\label{dirac-hamiltonian-unstrain}
H=-iv_{F}\int d^{2}x\left(\psi_{+}^{\dag}\vec{\sigma}\cdot\vec{\nabla}\psi_{+}-\psi_{-}^{\dag}\vec{\sigma}^{\ast}\cdot\vec{\nabla}\psi_{-}\right),
\end{equation}
where $\psi_{\pm}$ are two-component Dirac spinors, $\vec{\sigma}\equiv \left(\sigma_{1},\sigma_{2}\right)$, and $\vec{\sigma}^{\ast}\equiv \left(\sigma_{1},-\sigma_{2}\right)$, with $\sigma_i$ the usual Pauli matrices, all emerging from the honeycomb lattice structure of Fig.~\ref{honeycomb}.

For strained graphene, the slope of the Fermi cone is changed and the Dirac points are shifted \cite{olivia-leyva-naumis}, $\vec{k}_{\pm}^{SD} =\vec{k}_{\pm}^{D}+\vec{w}_{\pm}$, where $w_{\pm1}=\mp A_{1}$ and $w_{\pm2}=\mp A_{2}$, so that the Hamiltonian is now
\begin{equation}\label{dirac-hamiltonian-homogeneous-strain}
H=-iv_{F}\int d^{2}x\left(\psi_{+}^{\dag}\vec{\sigma}\cdot\left(\vec{\nabla}+i\vec{A}\right)\psi_{+}-\psi_{-}^{\dag}\vec{\sigma}^{\ast}\cdot\left(\vec{\nabla}-i\vec{A}\right)\psi_{-}\right),
\end{equation}
where $A_1 \sim (u_{11} - u_{22})$ and $A_2 \sim u_{12}$, or
\begin{equation}\label{latticegaugefield}
A^{i} = \epsilon^{ij} K^{jkl}u_{kl} \;,
\end{equation}
where $i,j,k,... = 1,2$. This is the usual pseudogauge field interpretation of the strain effect in the graphene sheet \cite{gauge_fields_graphene}. The third rank anisotropic tensor $K^{ijk}$, contracted with the strain tensor $u_{ij}$, is defined in Appendix B, and will be analyzed in the following, as this will help to clarify the peculiar (anisotropic) nature of this gauge field.

For homogeneous strain, the unstrained fermion $\psi_{0}$ is related to the strained one $\psi_{\pm}$ via a phase transformation
\begin{equation}\label{fermionic-transformation_homogeneous}
\psi_{\pm}=e^{i\vec{w_{\pm}}\cdot\vec{x}}\psi_{0\pm}.
\end{equation}
The modification of the Hamiltonian \eqref{dirac-hamiltonian-homogeneous-strain} in the presence of an inhomogeneous strain is computed in \cite{dejuan-sturla-vozmediano}. In this case \eqref{fermionic-transformation_homogeneous} becomes
\begin{equation}\label{formal_solution_strain}
\psi(\vec{x})=e^{i\int \vec{w}_{\pm}(\vec{x'})\cdot\vec{dx'}}\psi_{0}(\vec{x})=e^{\mp i\int \vec{A}(\vec{x'})\cdot\vec{dx'}}\psi_{0}(\vec{x}),
\end{equation}
where we follow the Dirac prescription  \cite{jackson-okun, berry1980}, $\psi_{0}$ is the unstrained solution. As a consequence of time-reversal symmetry, each Dirac point is charged with opposite sign. We can select a point $\vec{x}_{0}$ as a reference strain, i.e., a point where the strain effect $\vec{w}(\vec{x}_{0})$ could be gauged away. The phase acquired in \eqref{formal_solution_strain} could be interpreted as the circulation from the zero strain region $\vec{x}_{0}$ to the strained one $\vec{x}$. We stress on the fact that the solution \eqref{formal_solution_strain} is \emph{formal}: once the curl of $\vec{w}$ is non-zero, the integral in \eqref{formal_solution_strain} is path-dependent (is a non-reversible process). Once we realize that the effect of the inhomogeneous strain is to shift the momenta of the Fourier modes in a space-dependent way, we can interpret the situation as the $\psi$ fermion is in presence of an \emph{effective} gauge field $\vec{A}$, which could give us a non-zero effective magnetic field if the curl of $\vec{A}$ is non-zero, giving rise to the characteristic Landau levels. This gauge field, frequently called in the literature ``pseudo gauge field'', is the one we announced in the preceding part of this paper. Clearly, it could not have guessed from the QFT in curved space description.

We can follow the trail of the non-trivial behaviour of $\vec{A}$ as the result of the contraction of the strain tensor $u_{ij}$ with the third rank tensor $K^{ijk}$. This tensor is nonisotropic\footnote{In fact, the only isotropic tensors of rank three are proportional to the Levi-Civita antisymmetric tensor $\epsilon_{ijk}$, which is zero in two dimensions \cite{kearsley-fong}.} and its presence is due to the triad $\{\vec{s}_{1},\vec{s}_{2},{\vec{s}_{3}}\}$ specific to the structure of graphene, which is built from two sub-lattices \cite{gauge_fields_graphene}. The nontriviality resulting from this contraction could be seen in a simple example \cite{dejuan-manes-vozmediano}.

\begin{figure}
\begin{center}
\includegraphics[width=0.40\textwidth,angle=0]{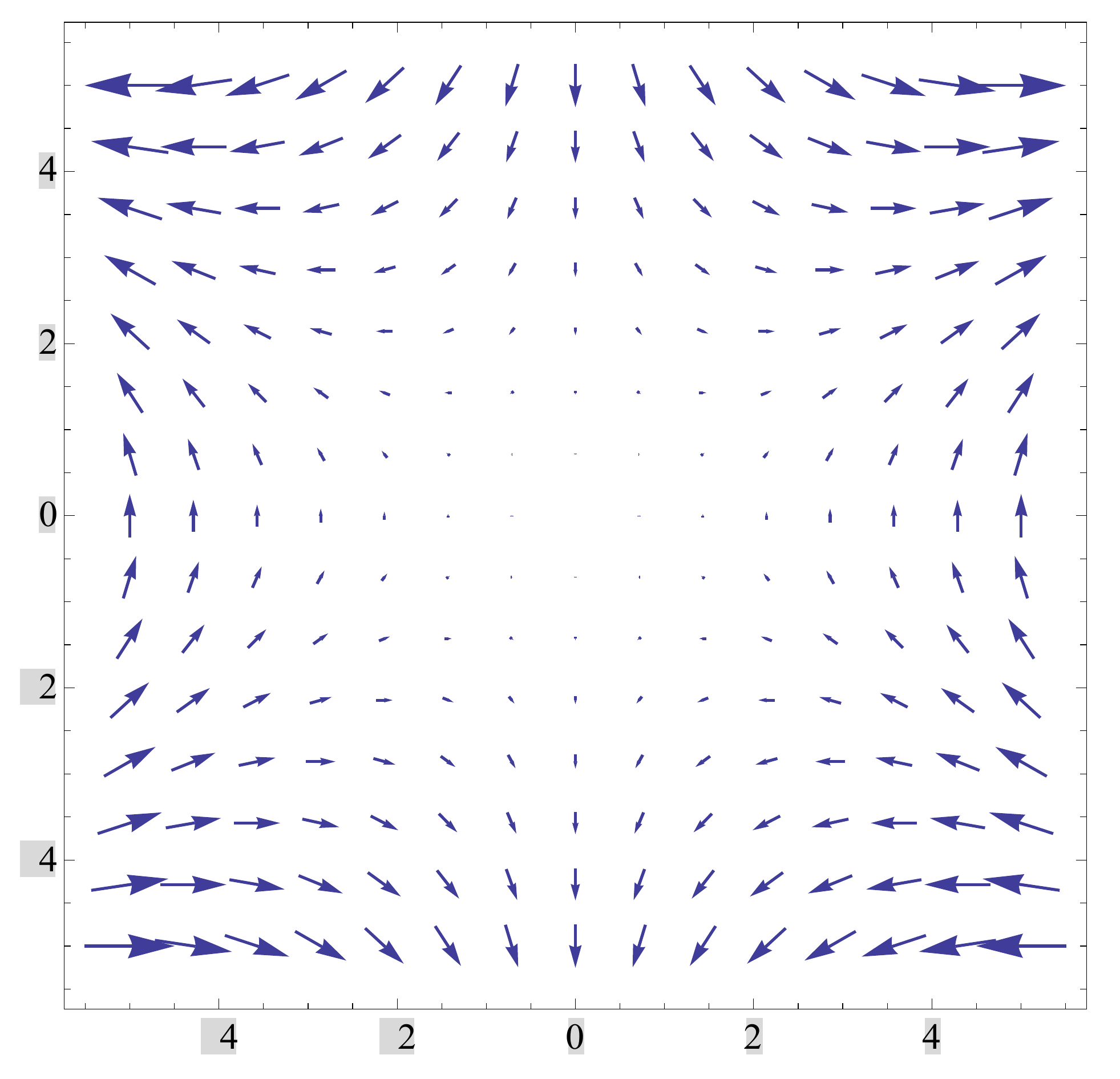}
\end{center}
\caption[3pt]{{\protect\small {Strained graphene with a deformation vector defined as $\vec{u}=(2xy,x^{2}-y^{2})u_{0}/4L$.}}}%
\label{strained-graphene}%
\end{figure}

\begin{figure}
\begin{center}
\includegraphics[width=0.40\textwidth,angle=0]{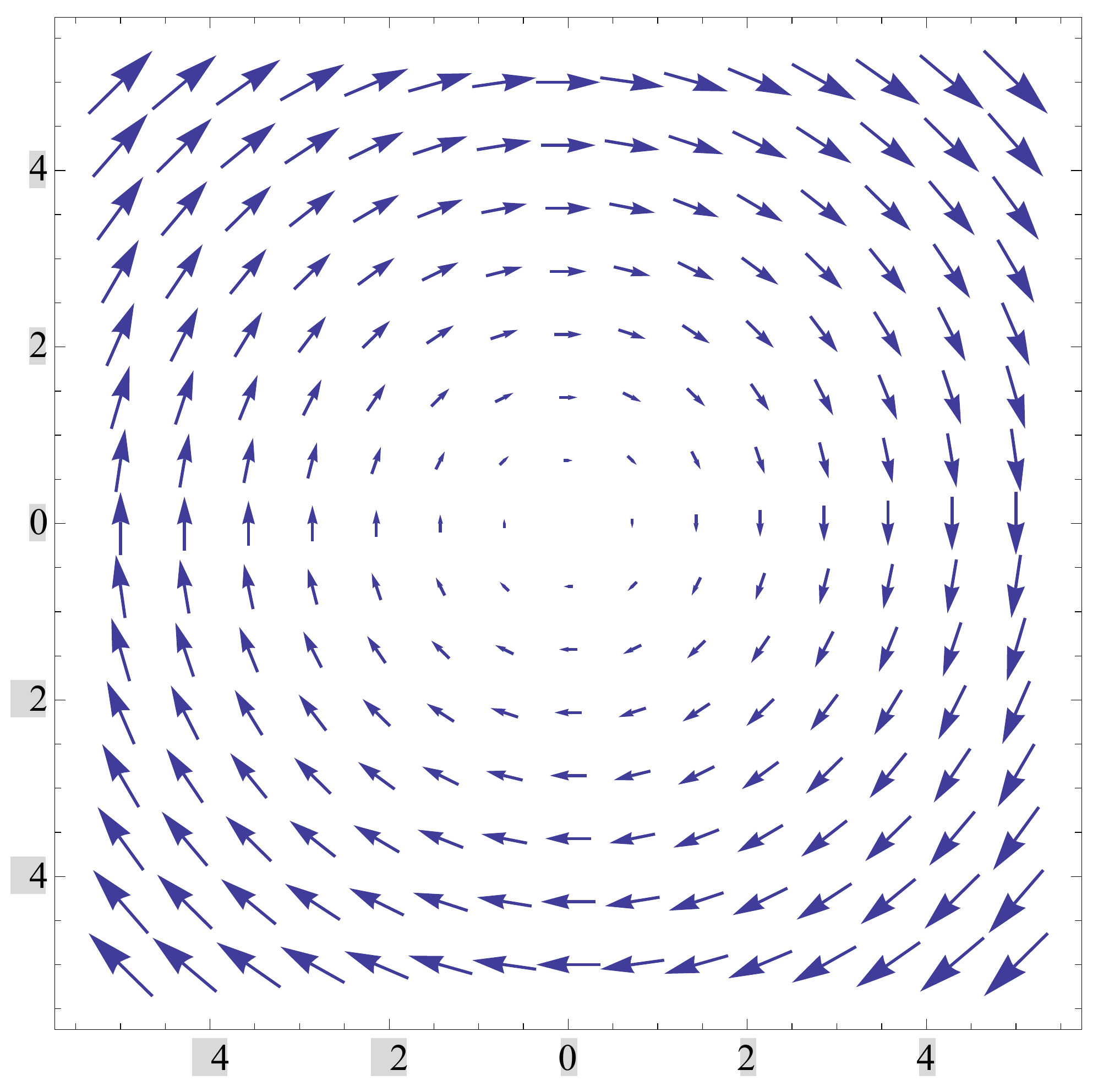}
\end{center}
\caption[3pt]{{\protect\small {Pseudogauge field produced using the deformation vector $\vec{u}=(2xy,x^{2}-y^{2})u_{0}/4L$.}}}%
\label{pseudo-gauge-field}%
\end{figure}

Consider the deformation vector $\vec{u}=(2xy,x^{2}-y^{2})u_{0}/L$, where $u_{0}$ is the maximum value of the strain and $L$ is the length of the graphene sample, as represented in Fig.~\ref{strained-graphene}.
This is a nonsingular vector field, in the sense that its curl and divergence is zero, so it is an irrotational vector field without any source or sink. However, we notice here that the vector field $\vec{A}$ resulting from the contraction of the corresponding strain tensor with $K^{ijk}$ is a clockwise rotational vector field with constant curl, as we can see in Fig.~\ref{pseudo-gauge-field}.
This means that not only does the displacement vector $\vec{u}$ matter but also the {\it orientation} of this vector with respect to the near-neighbor vector triad $\{\vec{s}_{i}\}$. So, the $K^{ijk}$ carries some memory of the lattice structure (short-distance behavior), even in the continuum limit (large-distance behavior).

\begin{figure}
\begin{center}
\includegraphics[width=0.70\textwidth,angle=0]{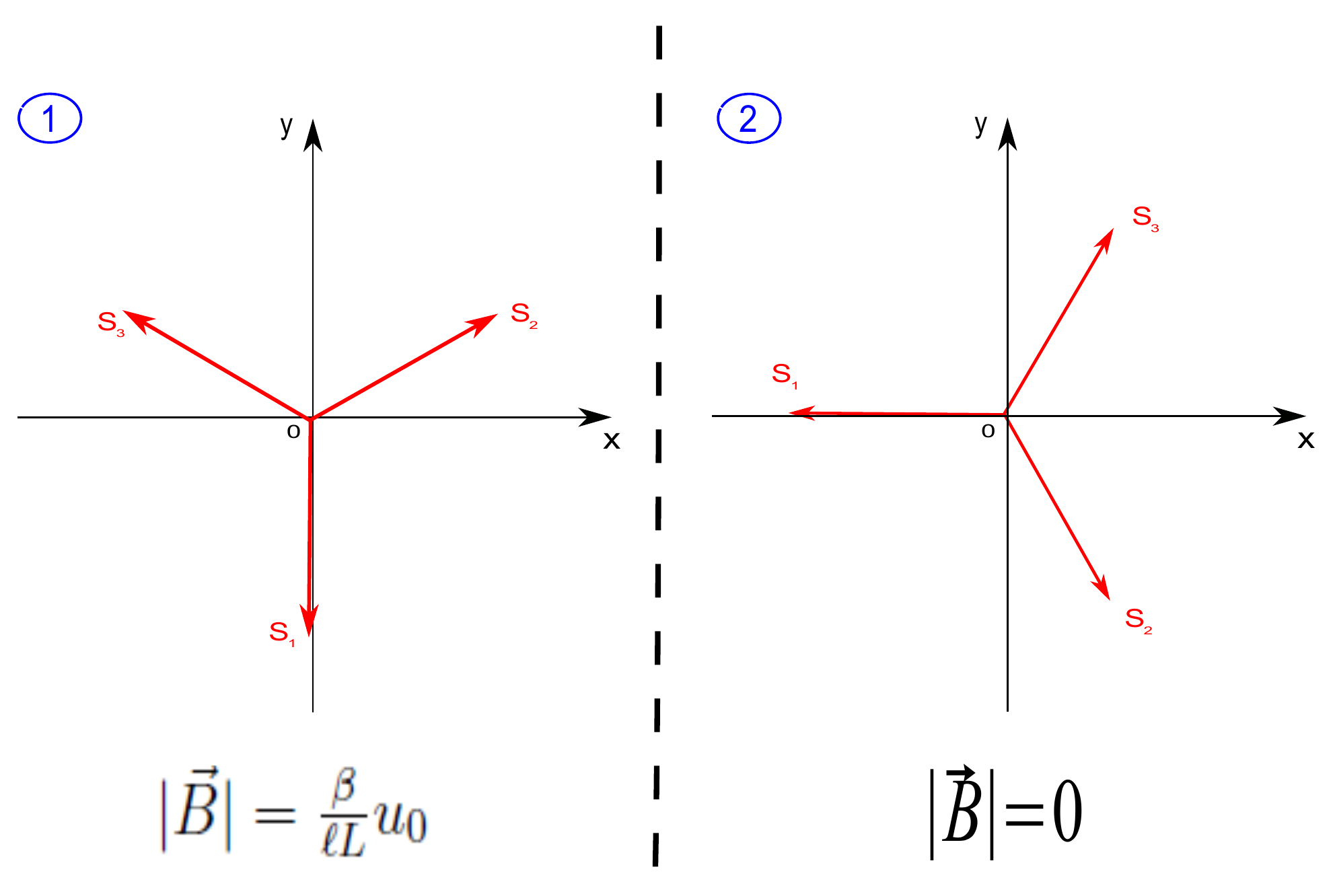}
\end{center}
\caption[3pt]{{\protect\small {Two samples of graphene unstrained sheets. Sample $2$ is clockwise rotated respect to sample $1$ by $\pi/2$ radians. Once we apply a deformation vector shown in Fig. \ref{strained-graphene}, the sample $1$ acquires a constant pseudo-magnetic field $|\vec{B}|=\frac{\beta}{\ell L} u_{0}$, while for the sample $2$ the pseudo-magnetic field is zero.}}}%
\label{graphene-samples}%
\end{figure}

To see more concretely that this is the case, suppose that we have two unstrained graphene sheet samples. The honeycomb orientation of both samples are such that one of them, say it sample $1$, has the vector $\vec{s}_{1}$ of the triad parallel to the $y$-axis, as in Fig.~\ref{honeycomb}, while sample $2$ has vector $\vec{s}_{1}=\ell(-1,0)$ parallel to the $x$-axis, see Fig. \ref{graphene-samples}. The orientation of the triad are both the same. A direct computation shows that the $K^{ijk}$ tensor are different for sample $1$ and sample $2$. Now let us apply to both samples the same strain deformation vector shown in Fig.~\ref{strained-graphene}. For the sample $1$, the pseudogauge vector field gives us a constant pseudomagnetic field $|\vec{B}|=\frac{\beta}{\ell L} u_{0}$, while for the sample $2$ the pseudo-magnetic field is zero. So, even though in the continuum limit the two samples only differ by their orientation, hence should be indistinguishable by isotropy, the pseudo-magnetic field keeps track of the honeycomb orientation at the lattice level. A detailed study of the so-called ``memory tensors'' in the graphene honeycomb and Kagom\'e lattices is found in \cite{cabra-grandi-silva-sturla}.

Once the curl of $\vec{w}$ is not zero, equation \eqref{formal_solution_strain} is not single-valued, as we can see if we take a loop around the origin. We are assuming that the variation of the strain tensor $u_{ij}$ is very small so, following the lines of \cite{berry1984,shapere-wilczek}, we can envisage a process to extract a physical meaning of the solution \eqref{formal_solution_strain}, even if it is not single-valued.

\begin{figure}
\begin{center}
\includegraphics[width=0.70\textwidth,angle=0]{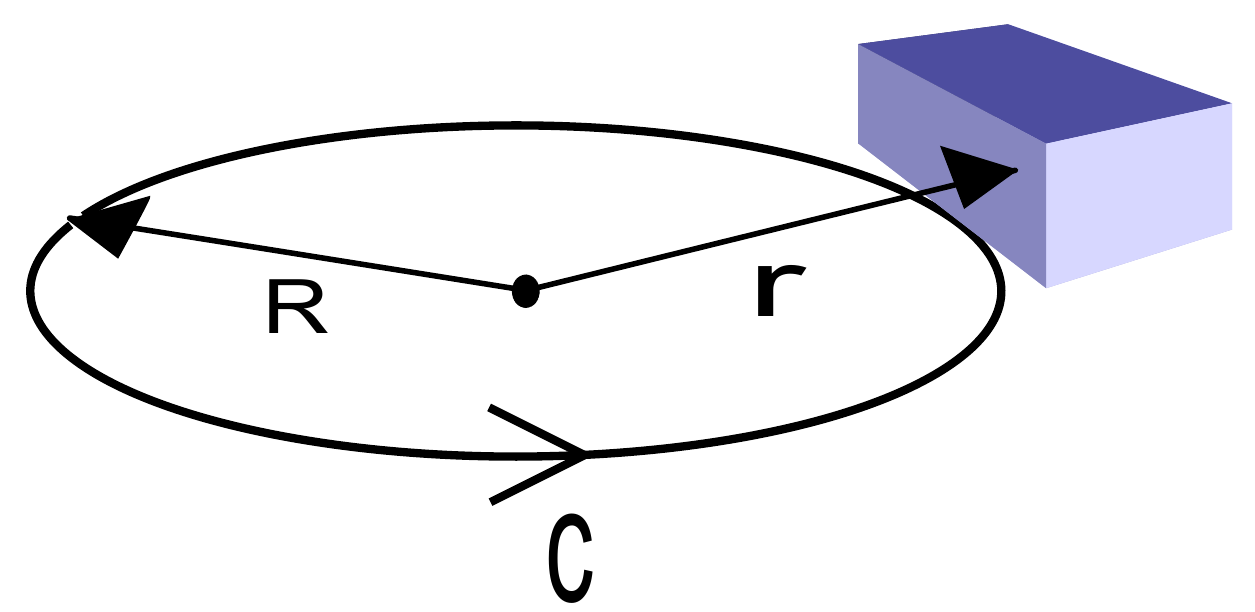}
\end{center}
\caption[3pt]{{\protect\small {Loop $C$ of the box (in blue) situated at $\vec{R}$. All the fermions inside the box could be described as $\psi(\vec{r}-\vec{R})$.}}}%
\label{geometric-phase}%
\end{figure}

Consider a small planar box on the graphene sheet (but very large compared with the lattice length $\ell$) situated at $\vec{R}$ (see Fig. \ref{geometric-phase}).
In the case of unstrained graphene ($\vec{A}=0$), the solutions for the $\pi$-electrons have the form\footnote{Because we are in the second quantization formalism, $\psi$ are operators and the wave packets are these operators applied to the vacuum.} $\psi_{0}^{\dag}(\vec{r}-\vec{R})=\psi_{0}|0>$. Now, the idea is to strain the graphene sheet in such a way that in the box the strain is homogeneous (constant shift $\vec{w}$ of the Dirac point as in the previous section) and the associated magnetic field is almost zero in that region. The solution of the $\pi$ electrons in the box could be written using the Dirac prescription \eqref{formal_solution_strain}
\begin{equation}\label{box_solution}
\psi(\vec{r}-\vec{R})=e^{- i\int_{\vec{R}}^{\vec{r}} \vec A(\vec{r'})\cdot\vec{dr'}}\psi_{0}(\vec{r}-\vec{R}).
\end{equation}
We can see that $\psi$ in \eqref{box_solution} is single valued in $\vec{r}$ and $\vec{R}$ locally. Now let the box be transported around a loop $C$. After completion of the loop, there will be a geometrical phase change that can be computed using Berry's formula \cite{nakahara}
\begin{equation}\label{berry-formula}
\gamma(C)=i\oint_{C}\psi^{\dag}(\vec{r}-\vec{R})\nabla_{\vec{R}}\psi(\vec{r}-\vec{R})\cdot d\vec{R}
\end{equation}
Taking into account that the $\psi$ are normalized, we end up with the result
\begin{equation}\label{strained-phase}
\gamma(C)= -\oint_{C} A(\vec{R})\cdot\vec{dR}=-\Phi,
\end{equation}
where $\Phi$ is the magnetic flux enclosed by the loop.

\section{Conclusions}

When only strain is present, the only gauge field that has unambiguous physical effects, is the one just discussed in Section \ref{bottom-up}.
The structure we saw there, is reminiscent of the gauge field arising in the Aharonov-Bohm effect, although strictly speaking, in order to have this effect, we need a magnetic flux line crossing the loop $C$ and a zero magnetic field outside the line. This is also reflected in the fact that the flux obtained in \eqref{strained-phase} is not constant and depends on the area enclosed by the loop. However, it could be possible to imagine a strain vector concentrated in some region of the graphene and assumed to be zero outside this region. Then, this procedure, recover the Aharonov-Bohm result. The Aharonov-Bohm effect is not new in the graphene literature and some examples of strain vectors were proposed in order to see this effect with particular procedures, see for instance \cite{chaves,ab-graphene}.

This gauge field could not have guessed from a top-down approach based on the standard QFT in curved spacetime. In fact, we saw here that if we define standard classical functions on graphene membrane the only fields emerging are the metric and the spin connection, which do not reproduce the behavior of this pseudogauge field. This $U(1)$ field can be put in correspondence with quantum field theoretical structures, such as the quantum anomalies. To see it, one just needs to realize that the origin of such gauge field is entirely quantum mechanical, and related to the fact that a ``translation'' in configuration space $T: \vec{x} \to \vec{x} + \vec{u}$, that is the straining of the graphene membrane, is necessarily associated with a ``translation''\footnote{A better name for this operation is ''Galilean boost'', that is why we use ``B'' for it. On this note see, e.g., \cite{weinberg}.} in momentum space $B: \vec{k} \to \vec{k} + \vec{v}$. Hence, in the first quantization language of the wave functions, those operations are carried on by quantum operators ${\cal U}: T\times B \to \mathbb{C}$ that, for the very meaning of quantum mechanics (that is the Heisenberg uncertainty principle) need to obey
\be\label{anomaly}
{\cal U}(T_1, B_1) {\cal U}(T_2, B_2) = e^{(\vec{u}_1 \vec{v}_2 - \vec{u}_2 \vec{v}_1)} {\cal U}(T_2, B_2) {\cal U}(T_1, B_1)  \;,
\ee
that are recognized, in the second quantization language, as instances of the nontriviality of the quantum field theoretical vacuum, and in turn of the quantum anomaly \cite{jackiw}.

Therefore, our simple table-top laboratory, can help explore yet another arena of fundamental physics, that is the deep meaning of the quantum anomaly\footnote{In our view, the very existence of a quantum anomaly is a sign of lack of understanding of how nature works at the most fundamental level, or, in other words, it is a sign of our ignorance of what is a proper quantum Noether theorem \cite{lop}.}.

On the other hand, this pseudogauge field, for the use of graphene we have in mind, is intriguing also because, as explained in the paper, it is a memory of the lattice structure (that is the physics of the wavelengths comparable to $\ell$, the lattice spacing) in the continuum limit (that is the physics of the large wavelengths). That means that it is a relic at low energy of the high energy behavior of the system. This is yet another reason why we cannot reproduce this field from a top-down approach, because in such approach the isotropy of the graphene membrane is tacitly assumed.

Effects of this kind would be of paramount importance to bring high energy theoretical constructions under the control of experiments. One example that comes to mind is the standard model extension of \cite{kostelecky}, where tensorial fields that are relics of the Lorentz invariant high energy string theory combine with the fields of the Standard Model (SM) and their derivatives within Lorentz violating terms that have the form
\begin{equation}\label{SME1}
T^{(k)}_{\;\;\;\;\mu ... \nu} ({\rm SM \; fields \; and \;
derivatives})^{\mu ... \nu} \;,
\end{equation}
where $T^{(k)} \sim \ell_{\rm Planck}^k$. In our ``graphene universe'' $\ell_{\rm Planck} \thickapprox \ell$.

We plan to explore these scenarios, and to address the full variety of possible gauge fields in graphene, namely those arising beyond the pure strain limitation, in later work.

\section*{Acknowledgments}

A.~I. acknowledges the Czech Science Foundation (GA\v{C}R), Contract No. 14-07983S, for support, and warmly thanks the Centro de Estudios Cient\'{\i}ficos (CECs) of Valdivia in Chile, for the kind hospitality while some of this work was carried on. P.~P. was partially supported from Fondecyt grant 1140155.


\appendix

\section{From $(2+1)$-dimensional Lagrangian to static two-dimensional Hamiltonian} \label{appendixa}

To be specific, let us focus on the case on negative constant Gaussian curvature, although similar formulae must hold the pure strain/fully flat case. The coordinates are
\bea
T & = & e^{t/r} \sqrt{e^{2 \sigma (\tilde{y})} + r^2} \\
X & = & e^{t/r} e^{\sigma (\tilde{y})} \cos \tilde{x} \\
Y & = & e^{t/r} e^{\sigma (\tilde{y})} \sin \tilde{x}
\eea
and
\[
\Sigma = - \frac{1}{2} \ln (T^2 - X^2 - Y^2) + \ln r = - t/r \;.
\]
To write the Hamiltonian in the $\tilde{q}$ coordinates, we have to consider the action \eqref{afinal}
\be
\mathcal{A} = i \int dT dX dY  e^{2 \Sigma} \left[ \bar{\psi} \gamma^0 (\partial_T + \Sigma_T) \psi
+ \bar{\psi} \gamma^1 (\partial_X + \Sigma_X) \psi +
\bar{\psi} \gamma^2 (\partial_Y + \Sigma_Y) \psi \right] \;.
\ee
We have to write this in the $\tilde{q}^\mu$ coordinates. We need the Jacobian
\be \label{jacobian}
\| \frac{\partial Q^\mu}{\partial \tilde{q}^\nu} \| =
- e^{3t/r} \; \frac{r \; e^{2 \sigma(\tilde{y})} \; \sigma_{\tilde{y}}(\tilde{y})}{\sqrt{e^{2\sigma(\tilde{y})} + r^2}}
\ee
the three terms
\bea
\Sigma_T & = & -\frac{T}{T^2 - X^2 - Y^2} = - \frac{e^{-t/r}}{r^2} \sqrt{e^{2\sigma(\tilde{y})} + r^2} \\
\Sigma_X & = & \frac{X}{T^2 - X^2 - Y^2} = \frac{e^{-t/r}}{r^2} e^{\sigma(\tilde{y})} \cos \tilde{x} \\
\Sigma_Y & = & \frac{Y}{T^2 - X^2 - Y^2} = \frac{e^{-t/r}}{r^2} e^{\sigma(\tilde{y})} \sin \tilde{x}
\eea
and to re-express the derivatives, e.g., $\partial_X = \tilde{x}_X \partial_{\tilde x} + \tilde{y}_X \partial_{\tilde y}$, etc., where, as usual, $\sigma_{\tilde{y}}(\tilde{y}) = \partial_{\tilde{y}} \sigma(\tilde{y})$, $\tilde{x}_X = \partial \tilde{x}/\partial X$, etc. Then we use
\be
\mathcal{A}=\int dtd\tilde{x}d\tilde{y}\mathcal{L}(\tilde{q})=\int dt d\tilde{x}d\tilde{y} \left[\left\|\frac{\partial Q}{\partial \tilde{q}}\right\|(\pi_{\psi}\partial_{T}\psi)(\tilde{q})-\mathcal{H}\right],
\ee
from which we can read off the Hamiltonian $H=\int d\tilde{x}d\tilde{y} \mathcal{H}(\tilde{q})$.
The final expression is
\bea
H & = & - i \hbar v_F \int d\tilde{x} d\tilde{y}
\left( - \frac{r \; e^{2 \sigma} \; \sigma_{\tilde{y}}}{\sqrt{e^{2\sigma} + r^2}} \right)
{\Big [} \psi^{\dag} \left( - \frac{\sqrt{e^{2\sigma} + r^2}}{r^2} \right) \psi \label{phi} \\
& + & \psi^\dagger \gamma^0 \gamma^1 {\Big (} - e^{-\sigma} \sin \tilde{x} \; \partial_{\tilde x}
+ \frac{e^{-\sigma}}{\sigma_{\tilde y}} \frac{e^{2 \sigma} + r^2}{r^2} \cos \tilde{x} \; \partial_{\tilde y}
+ \frac{e^{\sigma}}{r^2} \cos \tilde{x} {\Big )} \; \psi \label{A1} \\
& + & \psi^\dagger \gamma^0 \gamma^2 {\Big (} e^{-\sigma} \cos \tilde{x} \; \partial_{\tilde x}
+ \frac{e^{-\sigma}}{\sigma_{\tilde y}} \frac{e^{2 \sigma} + r^2}{r^2} \sin \tilde{x} \; \partial_{\tilde y}
+ \frac{e^{\sigma}}{r^2} \sin \tilde{x} {\Big )} \; \psi {\Big ]} \label{A2}
\eea
where we have reintroduced $\hbar$ and the Fermi velocity $v_F$. Note that nothing depends on $t$, as it must be.

The formula above for $H$ gives a field of the type $\psi^{\dagger}\psi$ in \eqref{phi}
\be
\phi = \frac{1}{r} \; e^{2 \sigma} \sigma_{\tilde y}\;,
\ee
the $A_{1}$ and $A_{2}$ fields in the non-derivative terms of \eqref{A1} and \eqref{A2} (here we identify $A_{1}$ and $A_{2}$ according the Pauli matrices $\tau^{1}$ and $\tau^{2}$)
\be
A_{1} = -\frac{1}{r} \; \frac{\; e^{3 \sigma} \; \sigma_{\tilde{y}}}{\sqrt{e^{2\sigma} + r^2}} \; \cos \tilde{x} \;, \;
A_{2} = -\frac{1}{r} \; \frac{\; e^{3 \sigma} \; \sigma_{\tilde{y}}}{\sqrt{e^{2\sigma} + r^2}} \; \sin \tilde{x} \;,
\ee
and the space-dependent, inhomogeneous Fermi velocity tensor
\be
\left[ v_F(\sigma(\tilde{y}))\right]_i^{\; \; \; \tilde j} = v_F \; \left(\begin{array}{cc} v_{1 \tilde{x}} & v_{1 \tilde{y}} \\
v_{2 \tilde{x}} & v_{2 \tilde{y}} \\ \end{array} \right)
\ee
with
\bea
v_{1 \tilde{x}} & = &  \frac{r \; e^{\sigma} \; \sigma_{\tilde{y}}}{\sqrt{e^{2\sigma} + r^2}} \; \sin \tilde{x} \\
v_{1 \tilde{y}} & = & -\frac{e^\sigma}{r} \; \sqrt{e^{2\sigma} + r^2} \; \cos \tilde{x} \\
v_{2 \tilde{x}} & = & -\frac{r \; e^{\sigma} \; \sigma_{\tilde{y}}}{\sqrt{e^{2\sigma} + r^2}} \; \cos \tilde{x} \\
v_{2 \tilde{y}} & = & -\frac{e^\sigma}{r} \; \sqrt{e^{2\sigma} + r^2} \; \sin \tilde{x}
\eea
Therefore, obtaining the expression \eqref{H2phiavf(x)} we have in the main text.

\section{Some details of the tight-binding computations} \label{strained-hamiltonian-derivation}

We start with the effective function of the hopping energy $t_{i}$ respect to intercarbon distance \cite{olivia-leyva-naumis},
\begin{equation}\label{hopping-energy-formula}
t_{i}=-\eta \exp\left[-\beta(\frac{\mid \vec{s'_{i}} \mid}{\ell}-1)\right],
\end{equation}
where $\eta\simeq2.8$ eV is the equilibrium hopping energy and $\vec{s'_{i}}$, with $i=1,2,3$, are the variation of the basis vectors $\vec{s_{i}}$. In order to deal with inhomogeneous strain, we consider the expansion of \eqref{hopping-energy-formula} up to the first derivative of the strain tensor $u_{ij}(\vec{x})$, i.e.,
\begin{equation}\label{hopping-energy-expansion}
t_{i}=\eta\left[1-\frac{\beta}{\ell^{2}}(s_{i})^{m}u_{mn}(s_{i})^{n}+\frac{\beta}{2\ell^{2}}(s_{i})^{k}(s_{i})^{m}(s_{i})^{n}\partial_{k}u_{mn}\right],
\end{equation}
where with some abuse of notation, $(s_{i})^{m}$ stands for the $m$ component of the vector $\vec{s_{i}}$, the indices $k,m,n$ are contracted (dummy indices) and $\partial_{k}\equiv\frac{\partial}{\partial x^{k}}$. The tight-binding Hamiltonian, in the second quantization formalism\footnote{If we use the second quantization formalism, some subtitles appear with respect to the vacuum where the creation and annihilation operators act. This subtitles will not be considerer here because the system is simple enough but, in the case of curved graphene, extra care must be taken. This is because the presence of defects on the honeycomb makes the vacuum nontrivial \cite{iorio}.}, could be written as
\begin{equation*}
H=-\sum\limits_{\vec{x}\in L_{A}}\sum\limits_{i=1}^{i=3} \left(a^{\dag}(\vec{x})t_{i}b(\vec{x}+\vec{s_{i}})+c.c.\right),
\end{equation*}
where $L_{a}$ stands for the sublattice $A$ (see Fig. \ref{honeycomb}) and, in order to have a Hermitian Hamiltonian, we added the complex conjugate of the first term. Using the following expressions for the Fourier transforms
\begin{equation*}
a(\vec{x})=\sum_{\vec{k}}e^{i\vec{k}\cdot\vec{x}}a_{\vec{k}} \hspace{.3cm}, \hspace{1cm} b(\vec{x})=\sum_{\vec{k}}e^{i\vec{k}\cdot\vec{x}}b_{\vec{k}} \hspace{.3cm}, \hspace{1cm} u^{m}(\vec{x})=\sum_{\vec{k}}e^{i\vec{k}\cdot\vec{x}}u_{\vec{k}}^{m} \hspace{.3cm},
\end{equation*}
we can write the Hamiltonian in the Fourier space as
\begin{equation}\label{hamiltonian-tb-bare}
H=-\eta\sum_{\vec{k},\vec{q}}\sum_{i=1}^{i=3} \left(b^{\dag}_{\vec{k}-\vec{q}},a^{\dag}_{\vec{k}}\right) \left(
                                                                                                           \begin{array}{cc}
                                                                                                             0 & e^{-i(\vec{k}-\vec{q})\cdot\vec{s_{i}}}T^{\dag}_{i,\vec{q}} \\
                                                                                                             e^{i(\vec{k}-\vec{q})\cdot\vec{s_{i}}}T_{i,\vec{q}} & 0 \\
                                                                                                           \end{array}
                                                                                                         \right)  \left(
                                                                                                                    \begin{array}{c}
                                                                                                                      b_{\vec{k}-\vec{q}} \\
                                                                                                                      a_{\vec{k}} \\
                                                                                                                    \end{array}
                                                                                                                  \right)
                                                                                                         ,
\end{equation}
where we used the symmetry property of $u_{ij}$ and defined $T_{i,\vec{q}}$ as
\begin{equation*}
T_{i,\vec{q}}=\delta(\vec{q})+i(s_{i})^{m}u_{m,\vec{q}}(s_{i})^{n}-(s_{i})^{j}(s_{i})^{m}(s_{i})^{n}q_{j}q_{m}u_{n,\vec{q}}.
\end{equation*}
Now, we expand the Hamiltonian around one Dirac point $K_{\pm}$. For the sake of simplicity, we expand around $\vec{K}_{+}=\left(\frac{4\pi}{3\sqrt{3}\ell},0\right)$, such that $\vec{k}=\vec{K_{+}}+\vec{p}$.
We can work a little bit more in the matrix content of \eqref{hamiltonian-tb-bare} as
\begin{equation*}
\left(
  \begin{array}{cc}
    0 & T^{\dag}_{i,\vec{q}}e^{-i(\vec{K_{+}}-\vec{q})\cdot\vec{s_{i}}} \\
    T_{i,\vec{q}}e^{i(\vec{K_{+}}-\vec{q})\cdot\vec{s_{i}}} & 0 \\
  \end{array}
\right)
=
\frac{i}{\ell}\vec{\sigma}\cdot\vec{s_{i}}\sigma_{3}\left(\mathbb{I} + i\sigma_{3}(\vec{p}-\vec{q})\right)\left(
  \begin{array}{cc}
    T^{\dag}_{i,\vec{q}} & 0 \\
    0 & T_{i,\vec{q}} \\
  \end{array}
\right),
\end{equation*}
where in the last equality we made use of the identity\footnote{In the case of expanding around the other inequivalent Dirac point $\vec{K}_{-}=\left(-\frac{4\pi}{3\sqrt{3}\ell},0\right)$, the formula is
$$\left(
  \begin{array}{cc}
    0 & e^{-i(\vec{K_{+}}-\vec{q})\cdot\vec{s_{i}}} \\
    e^{i(\vec{K_{+}}-\vec{q})\cdot\vec{s_{i}}} & 0 \\
  \end{array}
\right)
=
\frac{-i}{\ell}\vec{\sigma^{*}}\cdot\vec{s_{i}}\sigma_{3}$$.} \cite{dejuan-sturla-vozmediano}
\begin{equation*}
\left(
  \begin{array}{cc}
    0 & e^{-i(\vec{K_{+}}-\vec{q})\cdot\vec{s_{i}}} \\
    e^{i(\vec{K_{+}}-\vec{q})\cdot\vec{s_{i}}} & 0 \\
  \end{array}
\right)
=
\frac{i}{\ell}\vec{\sigma}\cdot\vec{s_{i}}\sigma_{3}.
\end{equation*}
At this point, it will be useful to show the following identities \cite{dejuan-sturla-vozmediano}
\begin{equation*}
\sum_{i=1}^{i=3}(s_{i})^{m}=0,
\end{equation*}
\begin{equation*}
\frac{1}{\ell^{2}}\sum_{i=1}^{i=3}(s_{i})^{m}(s_{i})^{n}=\frac{3}{2}\delta^{mn},
\end{equation*}
\begin{equation*}
\frac{1}{\ell^{3}}\sum_{i=1}^{i=3}(s_{i})^{j}(s_{i})^{m}(s_{i})^{n}=-\frac{3}{4}K^{jmn},
\end{equation*}
\begin{equation*}
\frac{1}{\ell^{4}}\sum_{i=1}^{i=3}(s_{i})^{j}(s_{i})^{m}(s_{i})^{n}(s_{i})^{r}=\frac{3}{8}\left(\delta^{jm}\delta^{nr}+\delta^{jn}\delta^{mr}+\delta^{jr}\delta^{mn}\right),
\end{equation*}
The tensor $K^{jmn}$, defined in the third identity, is an invariant under the discrete $C_{3}$ rotations. This tensor is very important in the discussion about the anisotropy of the strain-induced gauge field. For our choice of basis vectors $\{\vec{s_{i}}\}$, its only nonzero components are $K^{222}=-K^{122}=-K^{212}=-K^{221}=1$. We also note that the other three tensors are all isotropic. Making use of all this, doing some standard algebra and going back to the configurations space via the anti-Fourier transforms of $a$, $b$ and $u^{m}$, we end up with the following Hamiltonian for the inhomogeneous strain
\begin{equation*}
H=-i\sum_{\vec{x}} \psi_{+}^{\dag}\sigma^{j}\left(v_{jm}\partial_{m}+iv_{F}A^{j}-v_{F}\Gamma^{j}\right)\psi_{+},
\end{equation*}
where $v_{jm}=v_{F}\left(\delta_{jm}-\frac{\beta}{4}(u_{nn}\delta_{jm}+2u_{jm})\right)$ is the celebrated space-dependent Fermi velocity \cite{dejuan-sturla-vozmediano}, $A^{j}=\epsilon^{jp}K^{pmn}u_{mn}$ is the pseudogauge field, and $\Gamma^{j}=\frac{\beta}{4}\left(\partial_{m}u_{jm}+\frac{1}{2}\partial_{j}u_{mm}\right)$ is a connection-like coefficient. The Fermi velocity in unstrained graphene is $v_{F}=\frac{3}{2}\eta\ell$.

\end{document}